\documentclass[
aps,
nofootinbib,
prd,
superscriptaddress,
tightenlines,
notitlepage,
twocolumn,
showpacs,
floatfix
]{revtex4-1}

\usepackage{graphicx}
\usepackage{dcolumn}
\usepackage{bm}

\usepackage{orcidlink}
\usepackage{amsmath}
\usepackage{latexsym}
\usepackage{amsfonts}
\usepackage{graphicx}
\usepackage{mathrsfs}
\usepackage{epstopdf}
\usepackage{subfigure}
\usepackage[utf8]{inputenc}
\usepackage{CJK}
\usepackage{float}
\usepackage{xcolor}
\usepackage{color}
\usepackage{hyperref}
\usepackage{diagbox}
\usepackage{bm}
\usepackage{lipsum}
\usepackage[normalem]{ulem}
\hypersetup{
colorlinks=true,
linkcolor=red,
citecolor=blue,
}

\newcommand{\reply}[1]{#1}

\begin{document}

\title{Neutron stars in the bumblebee theory of gravity}

\author{Peixiang Ji}
\affiliation{Department of Astronomy, School of Physics, Peking University, Beijing 100871, China}
\affiliation{Kavli Institute for Astronomy and Astrophysics, Peking University, Beijing 100871, China}

\author{Zhuhai Li}
\affiliation{Department of Astronomy, School of Physics, Peking University, Beijing 100871, China}
\affiliation{Kavli Institute for Astronomy and Astrophysics, Peking University, Beijing 100871, China}

\author{Lirui Yang}
\affiliation{Cavendish Laboratory, JJ Thompson Avenue, Cambridge CB3 0HE, UK}

\author{Rui Xu}
\affiliation{Department of Astronomy, Tsinghua University, Beijing 100084, China}

\author{Zexin Hu}
\affiliation{Department of Astronomy, School of Physics, Peking University, Beijing 100871, China}
\affiliation{Kavli Institute for Astronomy and Astrophysics, Peking University, Beijing 100871, China}

\author{Lijing Shao}\email[Corresponding author: ]{lshao@pku.edu.cn}
\affiliation{Kavli Institute for Astronomy and Astrophysics, Peking University, Beijing 100871, China}
\affiliation{National Astronomical Observatories, Chinese Academy of Sciences, Beijing 100012, China}

\begin{abstract}
Recently, theoretical studies on the bumblebee gravity model, a
nonminimally-coupled vector-tensor theory that violates the Lorentz symmetry,
have flourished, with a simultaneous increase in the utilization of observations
to impose constraints.  The static spherical solutions of neutron stars (NSs) in
the bumblebee theory are calculated comprehensively in this work.  These
solutions with different coupling constants reveal a rich theoretical landscape
for NSs, including vectorized NSs and NSs with finite radii but divergent
masses.  With these solutions, preliminary constraints on the asymptotic vector
field values are obtained through restrictions on the stellar radius.
\end{abstract}

\maketitle

\section{Introduction}

Compact stars, generating extremely strong gravitational field, are among the
few environments where strong-field tests of gravity can be
conducted~\cite{LIGOScientific:2018dkp, Kramer:2021jcw, LIGOScientific:2021sio,
Shao:2022koz, EventHorizonTelescope:2022xqj, Shao:2022izp, 2024rpgt.book...61H,
Freire:2024adf}.  Mergering compact star binaries that emit strong gravitational
waves (GWs) have enhanced our understanding of dense matter and strong-field
gravity~\cite{LIGOScientific:2018dkp, LIGOScientific:2018hze,
LIGOScientific:2018cki}.  The study of gravitational astrophysics has entered a
more precise multi-messenger era, following the observations of GW170817 and
GRB170817A~\cite{LIGOScientific:2017vwq, LIGOScientific:2017zic,
Goldstein:2017mmi, LIGOScientific:2017ync}.  Studying neutron stars (NSs) in
modified gravity theories has a long history.  In early studies of NSs in
scalar-tensor theories~\cite{Damour:1993hw}, it was found that a NS could
acquire a scalar charge, leading to nonperturbative effects within the star.
These energetically favored scalarized NSs can spontaneously transform from
their corresponding  counterparts in general relativity (GR) through a phase
transition triggered by tachyonic instability~\cite{Damour:1996ke,
Sennett:2017lcx}, and stabilize due to nonlinear effects
ultimately~\cite{Doneva:2022ewd}.  This process, called ``spontaneous
scalarization", has been widely investigated in many generalized scalar-tensor
theories~\cite{Freire:2012mg, Yazadjiev:2016pcb, Ramazanoglu:2016kul,
Shao:2017gwu, Zhao:2022vig, Xu:2020vbs, Hu:2021tyw, Silva:2017uqg,
Doneva:2017bvd, Xu:2021kfh}.  In addition, the scalarized objects have garnered
significant attention because the weak-field regions as the asymptotic part of
their spacetimes do not violate current observations within the Solar
System~\cite{Will:2014kxa}.

In the past decade, there has been growing interest in studying the structure of
NSs within the framework of theories of gravity that include an additional
vector field.  A relatively simple theory with an extra vector field where the
NS structure was obtained is the Einstein-{\ae}ther
theory~\cite{Jacobson:2007veq, Yagi:2013qpa}.  In this theory, a dynamical unit
timelike vector field is coupled to gravity, thereby any solution of the vector
field defines the 4-velocity of a local ``preferred" frame at each spacetime
point, breaking the local Lorentz symmetry.  The maximal mass of a NS in the
Einstein-{\ae}ther theory  is smaller compared to what is predicted by
GR~\cite{Eling:2007xh}.  The discovery of a massive NS would constrain
parameters in the theory or even rule out this kind of vector {\ae}ther field.

However, general vector-tensor theories do not require a fixed magnitude for the
vector field.  For example, there are two simplest nonminimal couplings, 
\begin{align}\label{eq:vector:terms}
		\eta A^\mu A_\mu R,\quad \zeta A^\mu A^\nu R_{\mu\nu},
\end{align}
where $A^\mu$ is a vector field, in the Lagrangian of Hellings-Nordtvedt (HN)
theory~\cite{Hellings:1973zz}.  The static spherical NS solutions for the first
type of coupling were calculated by~\citet{Annulli:2019fzq}, and vectorized NS
solutions were found within a centain range of $\eta$.  However, such vectorized
spherical stars must arise out of nonlinear effects (such as selected initial
conditions) rather than from a linear mechanism originating from spherical GR
stars.  Additionally, some studies (see e.g.  Refs.~\cite{Silva:2021jya,
Demirboga:2021nrc}) pointed out that many theories of gravity involving
nonminimally coupled vector fields face a challenge where triggers of
spontaneous vectorization through tachyonic instability may also lead to ghost
instability.  Gradient instabilities, similar to the ghost instability, have
also been found in black hole (BH) solutions in the bumblebee theory
recently~\cite{Mai:2024lgk}.

The Standard Model Extension (SME) is an effective field theory that describes
various Lorentz symmetry breaking effects~\cite{Colladay:1996iz,
Colladay:1998fq, Kostelecky:2003fs}.  The bumblebee model can be treated as a
concrete example of the minimal gravitational SME, and its action
is~\cite{Kostelecky:2003fs, Bailey:2006fd}
\begin{eqnarray}\label{eq:action}
    S=&&\int d^4x\sqrt{-g}\left(\frac{R}{2\kappa}+\frac{\xi}{2\kappa}B^\mu B^\nu
    R_{\mu\nu}-\frac{1}{4}B^{\mu\nu}B_{\mu\nu}-V(\cdot) \right)\nonumber\\
    &&+S_\mathrm{m},
\end{eqnarray}
where $g$ is the determinant of the metric $g_{\mu\nu}$, the constant $\kappa
\equiv 8\pi G$ with $G$ being the gravitational constant, and $S_\mathrm{m}$
represents the action for matter fields.  Here, $B_\mu$ is the bumblebee field,
and the field strength tensor is $B_{\mu\nu}=\partial_\mu B_\nu-\partial_\nu
B_\mu$, similar to the electromagnetic field.

Unlike the HN theory, the bumblebee theory features a self-interaction
potential,
\begin{eqnarray}\label{eq:potential}
    V=V(B^\lambda B_\lambda\mp b^2).
\end{eqnarray}
For a stable vacuum of spacetime, we require that the potential $V$ is minimized
at $B_\mu=b_\mu$ and $b^\mu b_\mu=\pm b^2$.  Thus, the Lorentz symmetry is {\it
spontaneously} violated as the bumblebee field has a nonzero vacuum value
$\langle B_\mu\rangle=b_\mu$ with preferred spacetime directions.

There are two disparate approaches to solve the structure of astrophysical
objects in the bumblebee theory of gravity in literature.  One is to treat $b^2$
as a parameter in Eq.~\eqref{eq:potential} and solve for the dynamic field
$B_\mu$, where $b^2$ is usually set to be a constant to create a background
with uniform magnitude~\cite{Kostelecky:2003fs, Bailey:2006fd}.
\citet{Paramos:2014mda} considered this scenario with a harmonic oscillator
potential
\begin{eqnarray}\label{eq:hop}
    V=\frac{k}{2}\big(B^\lambda B_\lambda\mp b^2\big)^2,\quad k\neq0 \,,
\end{eqnarray}
where the time component of the vector field is set to zero ($B_t=0$) for
simplicity. \reply{Another more common consideration suggests that the potential $V$ is supposed to
be minimized at the vacuum expectation value of the bumblebee field $b_\mu$.}  One then has
\begin{eqnarray}\label{eq:dvdx0}
     \left.\frac{dV}{d(B^\lambda B_\lambda)}\right|_{B_\mu=b_\mu}=0.
\end{eqnarray}
\reply{The solution for astrophysical objects is based on the bumblebee field $B_\mu$ being frozen in its vacuum expectation value.
Therefore, the particular form of the potential driving its dynamics is irrelevant.
Different assumptions about the vacuum expectation configuration result in various forms of the function $b_\mu$, which in turn lead to the emergence of different astrophysical structures.}
Typically, the constant-magnitude~\cite{Casana:2017jkc,Neves:2024ggn}
and divergence-free~\cite{Bertolami:2005bh} conditions are applied.

\reply{In this work, we adopt the latter approach to solve for the static spherical NS solutions within the bumblebee theory. However, we do not impose any assumptions on the vacuum expectation value.
There are two reasons for this approach.
First, the assumption regarding the vacuum expectation value is not unique and lacks sufficient discussion about its validity.
Second, under a specific assumption, the $b_\mu$ typically does not satisfy the equations of motion for the vector field, as given in Eq.~\eqref{vecEoM}. This indicates that the vacuum solution does not stem from the spontaneous breaking of Lorentz symmetry at a certain energy scale.

One way to address this issue is to assume that the bumblebee field is coupled to the matter field, i.e. the Lagrangian of the matter field in Eq.~\eqref{eq:action} becomes a functional of the bumblebee field as well.
This coupling introduces an additional vector flux from the matter field, which provides the necessary degrees of freedom to satisfy the equations of motion for the bumblebee field.
Despite this, there are still some minor issues to be taken care of.
It remains an open question whether there is any coupling between the bumblebee field and the matter sector outside the star.
Additionally, the reasonableness of the assumption regarding the vacuum expectation value of the bumblebee field when matter is present inside the star is also uncertain. Therefore, instead of imposing an artificial assumption on $b_\mu$, we treat the vacuum background as a specific solution of the theory resulting from the spontaneous breaking of Lorentz symmetry.
In this case, $b_\mu$ simply satisfies the equation of motion for $B_\mu$ in the absence of coupling between the bumblebee field and the matter, provided that condition \eqref{eq:dvdx0} is satisfied.}
In this scenario, a series of studies on bumblebee BH solutions have recently been conducted~\cite{Xu:2022frb, Liang:2022gdk, Xu:2023xqh, Mai:2023ggs, Hu:2023vsg, Mai:2024lgk}.

\reply{For readers who are not familiar with the bumblebee theory, we strongly suggest viewing it simply as solving astrophysical objects in a vector-tensor theory, where the nonniminal coupling is given by the second type of Eq.~\eqref{eq:vector:terms}.
This is because, in the field equations, the potential and its first derivative terms are both zero for the background solution, and the equation of $b_\mu$ is indistinguishable from the equation of $A_\mu$.
The only difference is that $A_\mu$ is a general vector field, while $b_\mu$ is a background vector field that characterizes the breaking of Lorentz symmetry.
We notice that the equivalence no longer holds when perturbations are made to the bumblebee field.} 
Hence, our results are complementary to~\citet{Annulli:2019fzq}, in
which the authors solved the NS structure in the HN theory with the other coupling in Eq.~\eqref{eq:vector:terms}, i.e. $\eta A^\mu A_\mu R$.

The paper is organized as follows.  We start with setting up the equations and
classification of solutions in Sec.~\ref{sec:cla}, and a preliminary analysis of
solutions is shown in Sec.~\ref{sec:asy}.  Section~\ref{sec:for:and:par} introduces the
numerical method that we apply to solve field equations, and discusses the
parameter space for physical NS solutions.  The numerical results for all
possible coupling constant $\xi$ are displayed in Sec.~\ref{sec:num}.  Finally,
we summarize our findings and give a brief outlook for the directions worthy of
further study in Sec.~\ref{sec:sum}.  The appendices list equations and figures
to supplement the main text.

We use $(-,+,+,+)$ as the sign convention for the metric and assume $c=1$
throughout the paper.  We also apply the unit such that $\kappa=8\pi G=1$ to
avoid the repeated appearance in Sec.~\ref{sec:for:and:par}, as well as in
Appendices~\ref{appx:3} and \ref{appx:4}.  For quantities commonly encountered
in astrophysics, we express them in more easily understandable units.

\section{Equations For Static Spherical Solutions}

The action~\eqref{eq:action} for the bumblebee gravity model yields the field
equations for the tensor field and the vector field~\cite{Kostelecky:2003fs,
Bailey:2006fd, Xu:2022frb}
\begin{eqnarray}
    \mathscr{E}_{\mu\nu}\equiv G_{\mu\nu} - \kappa\big(T_{\mu\nu}^\mathrm{m} +
    T_{\mu\nu}^\mathrm{vec} + T_{\mu\nu}^\xi\big)=0,\label{gravEoM}\\
    \mathscr{E}^\mu\equiv\nabla_\nu B^{\mu\nu}-\frac{\xi}{\kappa} R^{\mu\nu}
    B_\nu+2B^\mu\frac{dV}{d(B^\lambda B_\lambda)}=0,\label{vecEoM}
\end{eqnarray}
where $T_{\mu\nu}^\mathrm{m}$ is the energy-momentum tensor for conventional
matter, while the contributions from the massive vector field and from
nonminimal coupling between gravity and the bumblebee field are
\begin{widetext}
\begin{eqnarray}
    T_{\mu\nu}^\mathrm{vec} & = & g^{\alpha\beta}B_{\mu\alpha} B_{\nu\beta} -
    g_{\mu\nu}\left(\frac14B^{\alpha\beta}B_{\alpha\beta} + V\right)+2B_\mu
    B_\nu\frac{dV}{d(B^\lambda B_\lambda)},\\
    T_{\mu\nu}^\xi&=&\frac{\xi}{2\kappa}\Big[g_{\mu\nu}(R_{\alpha\beta} -
    \nabla_\alpha\nabla_\beta)B^\alpha B^\beta +
    2\nabla_\alpha\nabla_{(\mu}(B_{\nu)}B^\alpha)-4B^\alpha
    B_{(\mu}R_{\nu)\alpha}-\square(B_\mu B_\nu)\Big].
\end{eqnarray}
\end{widetext}

\subsection{Two classes of solutions}\label{sec:cla}

For static and spherical solutions, we use the metric ansatz
\begin{equation}\label{eq:ansatz}
    ds^2=-e^{2\nu}dt^2+e^{2\mu}dr^2 + r^2(d\theta^2 + \sin^2\theta d\phi^2),
\end{equation}
where $\mu$ and $\nu$ are functions of the radial coordinate $r$.  Further, we
assume vanishing polar and azimuthal components of the bumblebee field, written
as $b_\mu=(b_t,b_r,0,0)$. From now on, the bumblebee fields refer to those
vacuum background configurations denoted as $b_\mu$.  Then, the field strength
tensor, defined as $b_{\mu\nu}=\partial_\mu b_\nu-\partial_\nu b_\mu$, takes the
form
\[b_{\mu\nu}=
    \begin{pmatrix}
    0 & -b_t' & 0 & 0\\
    b_t' & 0 & 0 & 0\\
    0 & 0 & 0 & 0\\
    0 & 0 & 0 & 0
    \end{pmatrix}.
\]
Here we assume that both $b_t$ and $b_r$ have only $r$ dependence, and the prime
denotes the derivative with respect to $r$.  The field equations, with the
ansatz of metric and vector field substituted, can be found in
Appendix~\ref{appx:1}.

With the help of the partial derivative formula for covariant divergence and
Eq.~\eqref{eq:dvdx0}, the components of Eq.~\eqref{vecEoM} may be written as
\begin{eqnarray}\label{vecEoM'}
    \mathscr{E}^{\bar\mu}=\frac{1}{\sqrt{-g}}\partial_r\big(\sqrt{-g}b^{\bar\mu
    r}\big)-\frac{\xi}{\kappa}R^{\bar\mu\bar\mu}b_{\bar\mu},
\end{eqnarray}
where there is no summation for the index $\bar\mu =t,r,\theta,\phi$.  It is
easy to find that $\mathscr{E}^\theta=\mathscr{E}^\phi=0$ automatically, and
$\mathscr{E}^t=0$ leads to a second-order ordinary differential equation of
$b_t$.  Due to the symmetry of $b^{\mu\nu}$, the component $\bar\mu=r$ gives
\begin{equation}\label{2cls}
    \xi R^{rr}b_r=0.
\end{equation}
Equation~\eqref{2cls} gives two classes of solutions
\begin{subequations}
    \begin{eqnarray}
    &\text{Class I:}\quad &b_r=0,\\
    &\text{Class II:}\quad &R_{rr}=0.
\end{eqnarray}
\end{subequations}
In the work on static spherical BH solutions of the bumblebee
gravity~\cite{Xu:2022frb}, Class I that restricts the vector field corresponds
to a family of vacuum solutions with 4 parameters, while Class II that restricts
the spacetime geometry corresponds to a family of vacuum solutions with 6
parameters.  In this paper, we only consider NS solutions of Class I.

\subsection{Approximate behaviors of the NS solutions}\label{sec:asy}

We expand the quantities near the center of a NS as,
\begin{subequations}
    \begin{eqnarray}
        &&\rho(r)=\sum_{n=0}^\infty\rho_nr^n,\\
        &&p(r)=\sum_{n=0}^\infty p_nr^n,\\
        &&\nu(r)=\sum_{n=0}^\infty\nu_nr^n,\\
        &&m(r)=\sum_{n=0}^\infty m_nr^n,\\
        &&b_t(r)=\sum_{n=0}^\infty b_nr^n,
\end{eqnarray}
\end{subequations}
where $\rho$ and $p$ refer to the mass density and pressure respectively, and
$m$ is related to $\mu$ by
\begin{eqnarray}
    e^{2\mu(r)}=\left(1-\frac{2Gm(r)}{r}\right)^{-1}.
\end{eqnarray}
After comparing equations of motion order by order, we have the non-vanishing
leading-order coefficients,
\begin{subequations}
    \begin{eqnarray}
    m_3&=&\frac{4\pi}{3}\frac{3p_0b_0^2(2\kappa-\xi)\xi +
    2e^{2\nu_0}\kappa\rho_0}{2e^{2\nu_0}
    \kappa+\xi(\xi-2\kappa)b_0^2},\label{eq:m3}\\
	b_2&=&\frac{\kappa}{6} \frac{e^{2\nu_0}(3p_0+\rho_0)}{2e^{2\nu_0} \kappa +
	\xi (\xi-2\kappa)b_0^2}\xi b_0,\label{eq:b2}\\
	\nu_2&=&\frac{\kappa^2}{6} \frac{e^{2\nu_0}(3p_0+\rho_0)}{2e^{2\nu_0} \kappa
	+ \xi(\xi-2\kappa) b_0^2},\label{eq:nu2}\\
	p_2&=&-\frac{\kappa^2}{6}\frac{e^{2\nu_0} (3p_0+\rho_0)
	(p_0+\rho_0)}{2e^{2\nu_0} \kappa + \xi(\xi-2\kappa)b_0^2}.\label{eq:p2}
    \end{eqnarray}    
\end{subequations}

All higher-order coefficients can be acquired from the central values $\rho_0$,
$p_0$, $\nu_0$ and $b_0$.  If $b_0=0$, then $b_t(r)=0$ since coefficients
$b_n\propto b_0$ for all $n>0$.  The remaining Eqs.~\eqref{eq:m3},
\eqref{eq:nu2}, and \eqref{eq:p2} become
\begin{eqnarray}\label{eq:grexp}
    &&m_3=\frac{4\pi}{3}\rho_0,\quad\nu_2 =
    \frac{2\pi}{3}G(3p_0+\rho_0),\nonumber\\
    &&p_2=-\frac{2\pi}{3}G(p_0+\rho_0)(3p_0+\rho_0),
\end{eqnarray}
which give the same NS structure as in GR.

The bumblebee model of gravity recovers the Einstein-Maxwell theory if
$V(\cdot)=0$ and $\xi=0$ are used.  In this situation, the equation for the
vector field has the solution
\begin{eqnarray}
    b_t(r)=b_0+\int_0^r\frac{q_0}{r'^2}e^{\mu(r')+\nu(r')}dr',
\end{eqnarray}
inside the star, where $q_0$ is an integral constant.  Since the bumblebee field
has no source if $\xi=0$, the possible solutions of $b_t'$ is either zero
($q_0=0$) or it represents the electric field generated by a point charge at the
origin ($q_0\neq 0$). The latter is excluded due to the convergence requirement
at the center, i.e. $b_t'(0)=b_1=0$.  Another particular value of the coupling
is $\xi=2\kappa$, in which case Eq.~\eqref{eq:grexp} is also satisfied with
\begin{eqnarray}\label{eq:expex}
    b_2=\frac16(3p_0+\rho_0)b_0.
\end{eqnarray}
In conclusion, the GR solution of spacetime is recovered if $\xi=0$ or
$\xi=2\kappa$, equipped with a constant or nontrivial vector field respectively.
Table~\ref{tab:togr} summarizes all conditions that give rise to NS solutions in
GR if $b_r=0$.

\begin{table}[t]
\caption{\label{tab:togr}
Summary of conditions that go back to NS and spacetime solutions in GR.}
\renewcommand\arraystretch{1.3}
\begin{ruledtabular}
\begin{tabular}{lll}
\textrm{Condition}&
\textrm{Coefficients}&
\textrm{Vector field}\\
\colrule
$b_0=0,\ \forall\xi$  & \textrm{Eq.~\eqref{eq:grexp}, } $b_2=0$ & \textrm{Vanishing} \\
$\xi=0,\ b_0\neq0$ & \textrm{Eq.~\eqref{eq:grexp}, } $b_2=0$ & \textrm{Constant} \\
$\xi=2\kappa,\ b_0\neq0$ & \textrm{Eq.~\eqref{eq:grexp}, Eq.~\eqref{eq:expex}} & \textrm{Nontrivial}\\
\end{tabular}
\end{ruledtabular}
\end{table}

The asymptotic expansion of variables at infinity can be written as
\begin{subequations}\label{eq:infexpand}
\begin{eqnarray}
    &&\nu(r)=\sum_{n=1}^\infty\frac{\nu_{-n}}{r^n},\\
    &&m(r)=M+\sum_{n=1}^\infty\frac{m_{-n}}{r^n},\label{eq:minf}\\
    &&b_t(r)=X+\frac{2\sqrt{\pi}Q}{r}+\sum_{n=2}^\infty\frac{b_{-n}}{r^n}.
\end{eqnarray}
\end{subequations}
Here, all the high-order coefficients in the summation only depend on the
Arnowitt-Deser-Misner (ADM) mass $M$, the background vector field $X$, and the
vector charge $Q$.  The factor $2\sqrt\pi$ has been chosen so that the
Reissner-Nordstr\"om metric is recovered for the vacuum solution when $\xi=0$.
The specific recurrence relations are displayed in Appendix~\ref{appx:2}.

Nevertheless, only two of the quantities in $\big\{M, X, Q\big\}$ are
independent, as all three depend on just two central variables (see
Sec.~\ref{sec:for:and:par}).  This can be clearly illustrated by the $\xi=2\kappa$
example whose exterior solution is the Schwarzschild spacetime, thereby one has
the relation $GMX+\sqrt{\pi}Q=0$.  We point out that all the BH solutions are
stealth Schwarzschild type if $\xi=2\kappa$~\cite{Xu:2022frb}, which indicates
that the Birkhoff theorem is applicable in this case (at least for the $b_r=0$
class), i.e. the exterior solution of a spherically symmetric star is indeed a
part of a BH solution of the theory.

\section{Formalism and Parameter Space}
\label{sec:for:and:par}

To solve the static spherical NSs in the bumblebee theory, we assume
$b_\mu=(b_t(r),0,0,0)$ and treat stars as perfect fluid equipped with an
equation of state (EOS), $p=p(\rho)$.  Therefore, the energy-momentum tensor of
a NS constituted by conventional matter is
\begin{eqnarray}
    T^\mathrm{m}_{\mu\nu}=(\rho+p)u_\mu u_\nu+pg_{\mu\nu},
\end{eqnarray}
where $u_\mu=(e^{\nu(r)},0,0,0)$ is the 4-velocity of matter, while $\rho(r)$
and $p(r)$ are energy density and pressure in a local inertial reference frame
comoving with matter.

The Bianchi identity, together with the equations of motion in
Appendix~\ref{appx:1}, gives the  modified Tolman-Oppenheimer-Volkoff (TOV)
equations
\begin{eqnarray}
	p'&=&-(p+\rho)\nu',\label{eq:TOVp}\\
	\mu'&=&-\frac{\mathscr D}{\mathscr C r},\\
	\nu''&=&\frac{\mathscr A}{r^2}+(\mu'-\nu')\left(\nu'+\frac1r\right),\\
	 b_t''&=&\left(\mu'+\nu'-\frac{2}{r}\right) b_t'\nonumber\\
    &&+\;\xi\left(\nu''+\nu'^2-\mu'\nu'+\frac{2}{r}\nu'\right) b_t,\label{eq:TOVbt}
\end{eqnarray}
where the prime denotes the derivative with respect to $r$, and the four
functions of $r$ are
\begin{subequations}
\begin{eqnarray}
    &&\mathscr A\equiv r^2 \left[pe^{2\mu}+\frac12 b_t'^2 - \xi  b_t\nu'( b_t'-
    b_t\nu')e^{-2\nu}\right],\quad\quad\quad\\
    &&\mathscr B\equiv \xi  b_t\big[( \xi-2) b_t-4( b_t'-
    b_t\nu')r\big]r\nu'+2r\nu'e^{2\nu}\nonumber\\
    &&\qquad\; -\;e^{2\mu+2\nu}(\rho+p) r^2+ \xi  b_t'^2r^2,\\
    &&\mathscr C\equiv 2e^{2\nu}+ \xi( \xi-2) b_t^2,\\
    &&\mathscr D\equiv \mathscr B+ \xi( \xi-2) b_t^2\mathscr A.
\end{eqnarray}    
\end{subequations}

\begin{table*}[t]
    \caption{\label{tab:paraspace}
    Requirements on $b_0$ for a given $\nu_0$.}
    \renewcommand\arraystretch{1.8}
    \begin{ruledtabular}
    \begin{tabular}{cccc}
    \multicolumn{1}{c}{Condition} & $\xi<0$ & $0<\xi<2\kappa$ & $\xi>2\kappa$\\
    \colrule
    $m_3>0$ & $b_0^2<\frac{2\rho_0}{3p_0}
    \frac{e^{2\nu_0}\kappa}{\xi(\xi-2\kappa)}$ & $b_0^2 <
    \frac{2e^{2\nu_0}\kappa}{\xi(2\kappa-\xi)}$ &
    $b_0^2<\frac{2\rho_0}{3p_0}\frac{e^{2\nu_0}\kappa}{\xi(\xi-2\kappa)}$\\
    $p_2<0$ & \textrm{any} $b_0$ &
    $b_0^2<\frac{2e^{2\nu_0}\kappa}{\xi(2\kappa-\xi)}$ & \textrm{any} $b_0$\\
    \textrm{No divergence\footnote{$b_0^\mathrm{max}$ represents the maximum
    value $b_0$ that can be attained. It certainly depends on $\xi$ and
    $\rho_c$, but it is hard to express analytically.  There is no limit on
    $b_0$ of singularity if a compact star is considered, i.e. for NSs with
    $R<20$ km, in the range $\xi>2\kappa$.}} & $b_0^2<(b_0^\mathrm{max})^2$ &
    $X^2<\frac{2\kappa}{\xi(2\kappa-\xi)}\Leftrightarrow
    b_0^2<(b_0^\mathrm{max})^2$ & \textrm{any} $b_0$\\
    \end{tabular}
    \end{ruledtabular}
\end{table*}

We start the integration of modified TOV equations at the center of the star
with boundary conditions\footnote{All the necessary first derivatives vanish due
to the constraints imposed by the equations of motion. An alternative form of
modified TOV equations, which makes this more apparent, is provided in
Appendix~\ref{appx:3}. This differs from scalar-tensor theories, where the first
derivatives are often artificially set to zero.}
\begin{eqnarray}
    &&\nu(0)=\nu_0,\quad p(0)=p_c,\quad \mu(0)=0,\nonumber\\
    &&\nu'(0)=0,\quad b_t(0)=b_0,\quad b_t'(0)=0,
\end{eqnarray}
where $p_c=p(\rho_c)$.  The integration terminates on the stellar surface $r=R$
where the pressure equals to zero.

To handle the situation at infinity, we consider the coordinate transformation
\begin{eqnarray}
    x=\frac{r-R}{r+\beta R},
\end{eqnarray}
which maps the radius of the star to $x=0$ and maps the infinity to $x=1$.
Here, $\beta\geq0$ is a free parameter which controls the asymptotic behaviors
of variables that are treated as functions of $x$.  The modified TOV equations
outside the star as functions of $x$ are shown in Appendix~\ref{appx:4}.
Fortunately, $x=1$ is not a singularity of the differential equations, thereby
the integration outside a NS can be calculated.

The radial component of the metric and the time component of the vector field
can be expressed at infinity as
\begin{eqnarray}
    g^{rr}(r)&=&1-\frac{m(r)}{4\pi r},\label{eq:infgtt}\\
    b_t(r)&=&X+\frac{q(r)}{r},\label{eq:infbt}
\end{eqnarray}
where $m(r)$ is given in Eq.~\eqref{eq:minf} and
\begin{eqnarray}
    q(r)=2\sqrt{\pi}Q+\sum_{n=2}^\infty\frac{b_{-n}}{r^{n-1}}.
\end{eqnarray}
Differentiating Eq.~\eqref{eq:infgtt} and Eq.~\eqref{eq:infbt} with respect to $x$,
one finds
\begin{eqnarray}
    &&\dot\mu=\frac{(1+\beta x)(1-x)\dot m-(1+\beta) m}{4\pi(1+\beta
    x)^2R}e^{2\mu},\label{eq:infgttd}\\
    &&\dot b_t=\frac{(1+\beta x)(1-x)\dot q-(1+\beta)q}{(1+\beta x)^2 R},\label{eq:infbtd}
\end{eqnarray}
where the overdot denotes the derivative with respect to $x$.  The ADM mass and
the vector charge are then acquired by considering Eq.~\eqref{eq:infgttd} and
Eq.~\eqref{eq:infbtd} at $x=1$ respectively,
\begin{eqnarray}
    &&M=m\big|_{x=1}=8\pi(1+\beta)R\dot\nu \big|_{x=1},\\
    &&Q=\frac{q \big|_{x=1}}{2\sqrt{\pi}}=-\frac{1+\beta}{2\sqrt{\pi}}R\dot b_t \big|_{x=1},
\end{eqnarray}
where $(\dot\mu+\dot\nu)_{x=1}=0$ is used.  One can define the Schwarzschild
radius $R_\mathrm{S}\equiv M/4\pi$.  The stellar radius becomes smaller than its
Schwarzschild radius (see e.g. Ref.~\cite{Li:2023vbo}) if
\begin{eqnarray}
    2(1+\beta)\dot\nu \big|_{x=1}>1.
\end{eqnarray}

The modified TOV equations only guarantee $g_{rr}=1$ at infinity, and one needs
to choose the central values $p_c$, $\nu_0$, and $b_0$ carefully to ensure
$g_{tt}=-1$.  In other words, the requirement of asymptotic flatness makes only
two out of these three boundary variables $\big\{ p_c, \nu_0, b_0 \big\}$
independent.  To satisfy this, let us consider a coordinate transformation
\begin{eqnarray*}
    t\to e^{-\Delta\nu}t,
\end{eqnarray*}
while $r$, $\theta$ and $\phi$ remain the same.  The time components of the
metric $g_{\mu\nu}$ and the vector field $b_\mu$ change correspondingly,
\begin{eqnarray*}
    g_{tt}\to e^{2\Delta\nu}g_{tt},\quad b_t\to e^{\Delta\nu}b_t.
\end{eqnarray*}
Therefore, the system of Eqs.~(\ref{eq:TOVp}--\ref{eq:TOVbt}) is invariant under
the transformation
\begin{eqnarray}
    \nu\to\nu+\Delta\nu,\quad b_t\to e^{\Delta\nu}b_t
\end{eqnarray}
for an arbitrary $\Delta\nu$, and the asymptotic flatness is easily satisfied
after setting $\Delta\nu = -\nu_\infty$.  Consequently, the boundary conditions
switch to
\begin{eqnarray}
    \nu_0\to\nu_c=\nu_0+\Delta\nu,\quad b_0\to b_c=e^{\Delta\nu}b_0 .
\end{eqnarray}
The radius, ADM mass and vector charge of the NS, as well as the background
vector field at infinity, change correspondingly as
\begin{eqnarray}
    R\to R,\quad M\to M,\quad Q\to e^{\Delta\nu}Q,\quad X\to e^{\Delta\nu}X.
\end{eqnarray}

Generally, the vector field at infinity is determined by the background cosmic
evolution in the bumblebee theory.  It is natural to assume that all the NS
spacetime share the same asymptotic vector field, $\bar b_\mu=(\bar b_t,0,0,0)$.
Once the central density $\rho_c$ and the background vector field $\bar b_t$ are
chosen, the boundary condition $(p_c,\nu_c,b_c)$ is determined, which leads to a
unique asymptotically flat solution of a NS in the bumblebee theory with a given
coupling constant $\xi$.

A positive $m_3$ and a negative $p_2$ in Eq.~\eqref{eq:grexp}, which are
guaranteed by a series of energy conditions, lead to a gravitational mass that
increases with radius (i.e. $dm/dr>0$) and a pressure that decreases with radius
(i.e. $dp/dr<0$) in GR.  As an analogy, we display the requirements of
parameters in the bumblebee theory in Table~\ref{tab:paraspace}, which are
obtained assuming $m_3>0$ in Eq.~\eqref{eq:m3} and $p_2<0$ in Eq.~\eqref{eq:p2}.
On the one hand, $m_3$ could be negative for all $\xi$ except 0 and $2\kappa$,
which means a negative gravitational mass near the center.  A range of negative
gravitational masses indicates that the vector field, as a portion of the
gravitational force, exerts repulsion force, which might relate to the instability
of stars.  On the other hand, $p_2$ can be positive and
leads to an increasing pressure along the radius when $0<\xi<2\kappa$, which is
considered unphysical for a relativistic star.  Divergences frequently arise
during the process of solving the modified TOV equations, which also constrain
the values of variables at the stellar center. These are summarized in the last
row of Table~\ref{tab:paraspace}, with specific details on the divergences
discussed in Sec.~\ref{sec:num}.  At last, we require that the radius of the
star does not exceed 20 km, as we are interested in compact stars.

\section{Numerical results}\label{sec:num}

To numerically calculate the NS spacetime, we choose the AP4 EOS as an example.
There is no qualitative difference between different EOSs for NSs in the bumble
gravity theory.  Solutions are separated into three different parts by two
special values, namely $\xi=0$ and $\xi=2\kappa$, and each of those is discussed
in this section.  Without loss of generality, we set $\nu_0=0$ and assume
$b_0>0$, numerically calculating NS solutions with all possible $\rho_c$ and
$b_0$.  Some relevant quantities of  NSs are computed, including the baryonic
mass $M_{\rm b}$,
\begin{align}
	M_\mathrm{b}=m_\mathrm{b}\int
	n_\mathrm{b}\sqrt{-g}u^0d^3x=\int\rho_\mathrm{b}e^\mu d^3x,
\end{align}
 and three dimensionless quantities, compactness $C$, fractional binding energy
 $f_{\rm b}$, and charge-to-mass ratio $\alpha$, defined
 as~\cite{Misner:1973prb},
\begin{eqnarray}
    C\equiv\frac{GM}{R},\quad
    f_\mathrm{b}\equiv\frac{M_\mathrm{b}}{M}-1, \quad
    \alpha\equiv\frac{Q}{\sqrt{\kappa}M} .
\end{eqnarray}
Results are given below and in Appendix~\ref{appx:5}.

\subsection{The $\xi<0$ case}

\begin{figure}[t]
\includegraphics[width=\linewidth]{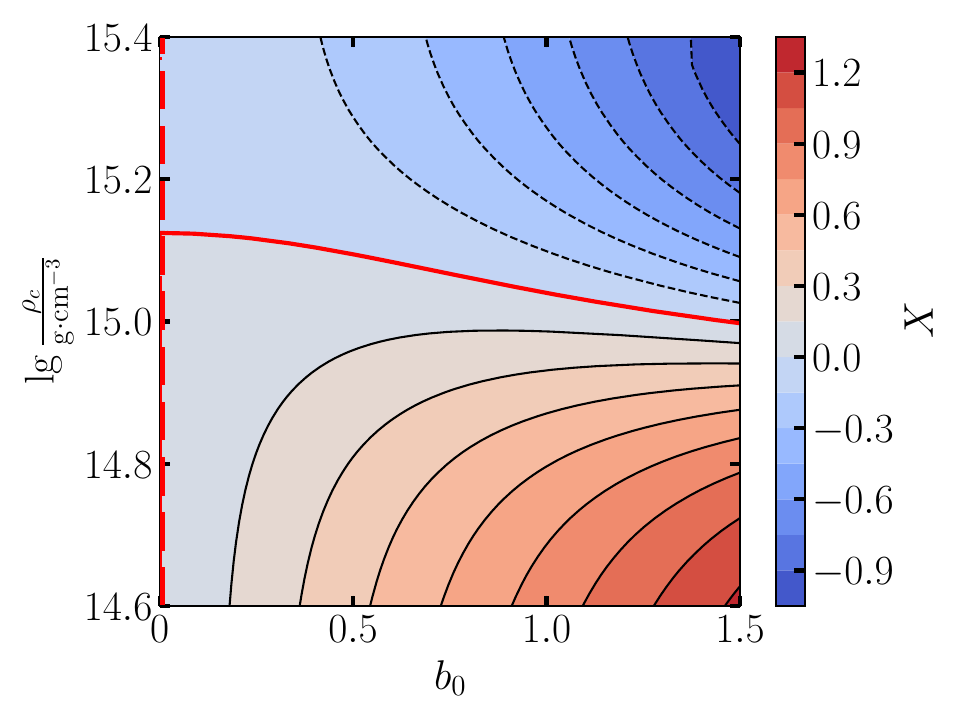}
\caption{\label{fig:contour} Contour plot for the vector field at infinity
[namely, $X$ in Eq.~(\ref{eq:infbt})] on the $b_0$-$\rho_c$ plane for
$\xi=-\kappa$.  The black solid lines, red solid lines, and black dashed lines
denote contours where $X<0$, $X=0$, and $X>0$, respectively.  The vertical red
dashed line corresponding to $b_0=0$ represents the GR solutions, and the red
solid curved line represents the vectorized solutions.}
\end{figure}

In Fig.~\ref{fig:contour}, we present the contour plot of the time component of
the bumblebee field at infinity for $\xi=-\kappa$ as an illustrative case for a
negative coupling constant.  When the boundary condition $b_0=0$ is imposed, the
modified TOV equations ensure that $b_t(r)=0$ throughout the spacetime, reducing
the NS solutions to those found in GR, as indicated by the vertical red dashed
line.  The red solid curve represents solutions that exhibit the same asymptotic
behavior as that in GR, including a cosmic background with $\bar b_t=0$, but
with a nontrivial vector field.  These vectorized solutions are only found
within a specific parameter space where $\xi<0$.

\begin{figure}[t]
\includegraphics[width=\linewidth]{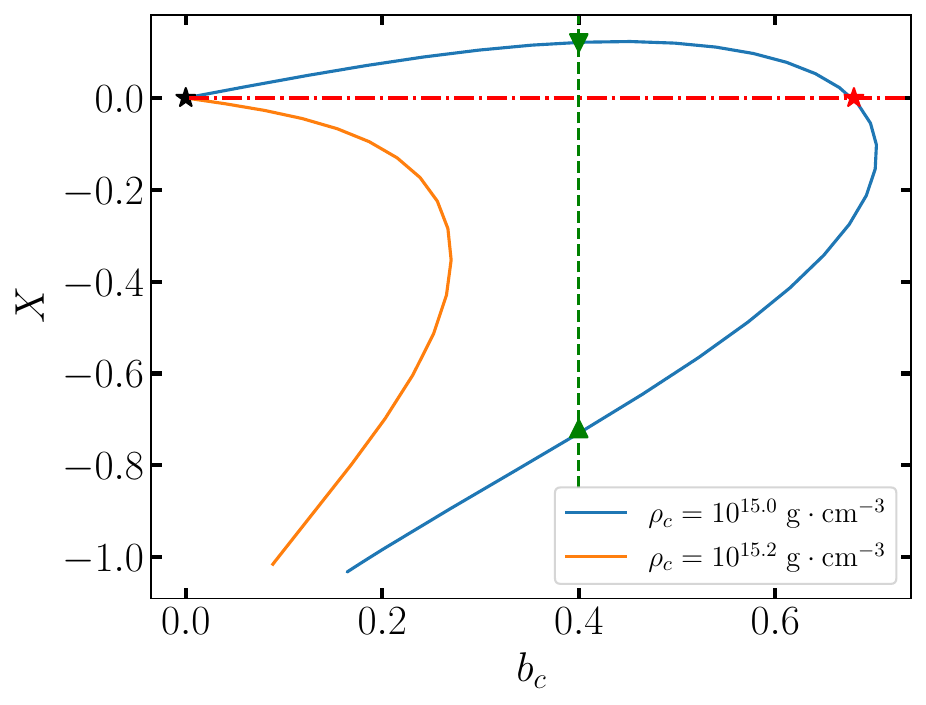}
\caption{\label{fig:b0-X} The time component of vector field at infinity as a
function of its central value of the star for $\xi=-\kappa$.  The blue and
orange curves represent two different central densities, and both of them
terminate because of the divergence of $g_{rr}$. The black and red stars
respectively denote a NS in GR and a vectorized NS, while the green triangle are
NSs sharing the same vector field at stellar center.}
\end{figure}

In Fig.~\ref{fig:b0-X}, we plot the relation between the vector fields at the
center and at infinity. The blue and orange curves correspond to
$\rho_c=10^{15.0}\;\mathrm{g\cdot cm^{-3}}$ and
$\rho_c=10^{15.2}\;\mathrm{g\cdot cm^{-3}}$, respectively. Any point on these
curves which intersect the red dot-dashed line  represents a vectorized NS,
denoted by a red star, in contrast to a NS solution in GR, which is given by a
black star.  It is apparent that as the central density increases, the
vectorized solution no longer exists.  Therefore, there is an upper bound for
$\rho_c$ for vectorized solutions, which can be determined from the intersection
of the two red lines in Fig.~\ref{fig:contour}.  While $\rho_c$ does not
theoretically have a lower bound, it cannot be too small given that we are
considering compact stars and the radius should be less than 20 km.

Another interesting feature of the bumblebee theory is that it allows for two
distinct asymptotically flat solutions with a same central density and vector
field, as exemplified by the green  triangles where
$\rho_c=10^{15}\;\mathrm{g\cdot cm^{-3}}$ and $b_c = 0.4$ in
Fig.~\ref{fig:b0-X}.  The difference of boundary condition between these two
solutions arises from the difference in the metric component $g_{tt}$ at the
center, more specifically from the difference in $\Delta\nu$, which is used to
set the boundary conditions at infinity.

\begin{figure}[t]
\includegraphics[width=\linewidth]{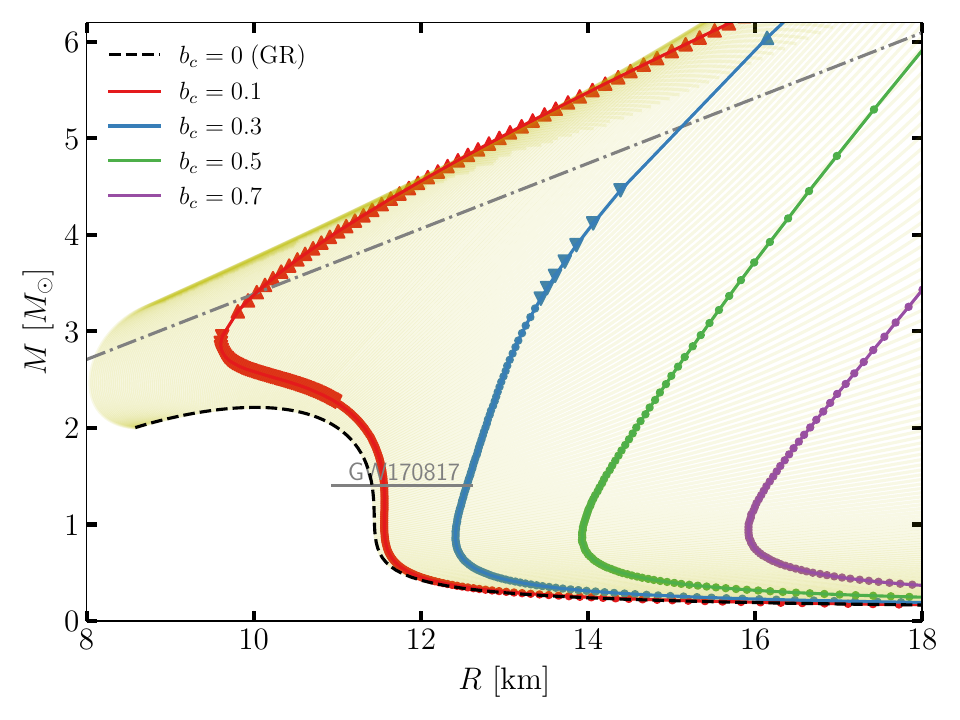}
\includegraphics[width=\linewidth]{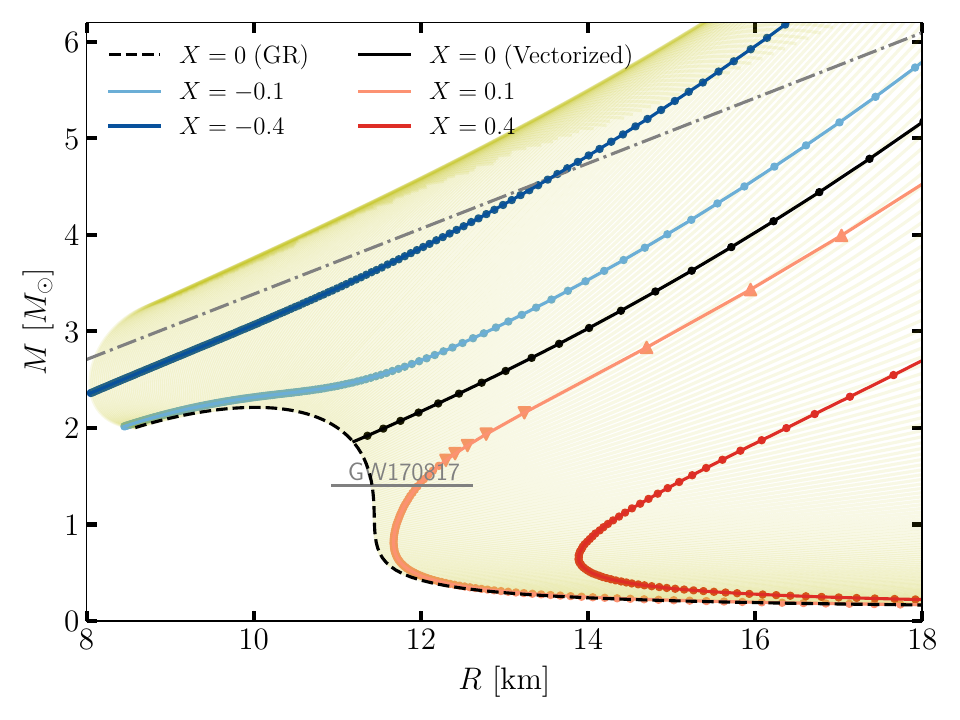}
\caption{\label{fig:-1MR} The mass-radius curves of NSs for $\xi=-\kappa$ with a
same vector field at the center (top panel) and at infinity (bottom panel). The
buff region represents all possible NSs in this case. In both panels, the black
dashed line is the mass-radius curve of NSs in GR, the gray dot-dashed line
denotes NSs with a compactness of 0.5, and the short gray bar marks the estimate
of the radius of a $M=1.4\,M_\odot$ NS~\cite{Dietrich:2020efo}.}
\end{figure}

The mass-radius relation of NSs in the $\xi=-\kappa$ bumblebee theory is shown
in Fig.~\ref{fig:-1MR}.  The buff region, which is actually composed of numerous
curves representing NSs with the same central density when viewed closely, marks
the mass and radius of all possible NSs in this scenario.  Each curve in the top
(bottom) panel represents NSs that share the same vector field at the stellar
center (at infinity). All dots correspond to specific stars, with different
markers indicating different characteristics.  If there is only one solution for
a fixed central density $\rho_c$ that satisfies the set of boundary conditions
(a fixed $b_c$ or $X$), the corresponding NS is represented by a general point.
However, if there are two solutions for a fixed $\rho_c$, such as the two green
triangles in Fig.~\ref{fig:b0-X}, they are labeled as triangular points.
Clearly, an upward triangular point and a downward triangular point always
appear in pairs, where the upward (downward) triangular point represents the NS
with a larger (smaller) fractional binding energy $f_\mathrm{b}$.  The closest points on a
curve form a pair of upward and downward triangular points, and additional pairs
can be identified by removing known pairs and repeating the process.  It is easy
to imagine that there must be a point between the nearest pair of triangular
points where the upward triangle  and the downward triangle coincide, referring
to the tangent points (the rightmost point for the same $b_c$ case while the
upmost point for the same $X$ case) of the $b_c$-$X$ curve with the horizontal
or vertical line in Fig.~\ref{fig:b0-X}.

The black dashed line in Fig.~\ref{fig:-1MR} represents the mass-radius curve of
a NS in GR.  The black solid line, which represents vectorized NSs, bifurcates
from this curve.  We find that the vectorized NS has a larger binding energy
compared to the NS in GR (explicitly illustrated in Fig.~\ref{fig:appx:1} in
Appendix~\ref{appx:5}) with the same central density, suggesting that vectorized
stars are more energetically favorable.  As the coupling becomes more negative,
the corresponding bifurcation point has a lower central density, resulting in a
shift toward the lower right of the mass-radius curve in GR. This result is
quite similar to the findings by~\citet{Annulli:2019fzq} studying another form
of nonminimal coupling $\eta A^\mu A_\mu R$ in Eq.~\eqref{eq:vector:terms}, which suggests a certain generality
within vector-tensor theories.

The short gray bar indicates the estimate from GW170817
\cite{LIGOScientific:2017zic} for the radius of a $M=1.4M_\odot$ NS, with a 90\%
confidence level~\cite{Dietrich:2020efo}.  Given an EOS, the estimated value,
$R=11.75^{+0.86}_{-0.81}$ km, can be used to constrain the vector field at
infinity.  Specifically, for the coupling $\xi=-\kappa$, the constraint is
$X\lesssim0.17$.

\subsection{The $0<\xi<2\kappa$ case}

\begin{figure}[t]
\includegraphics[width=\linewidth]{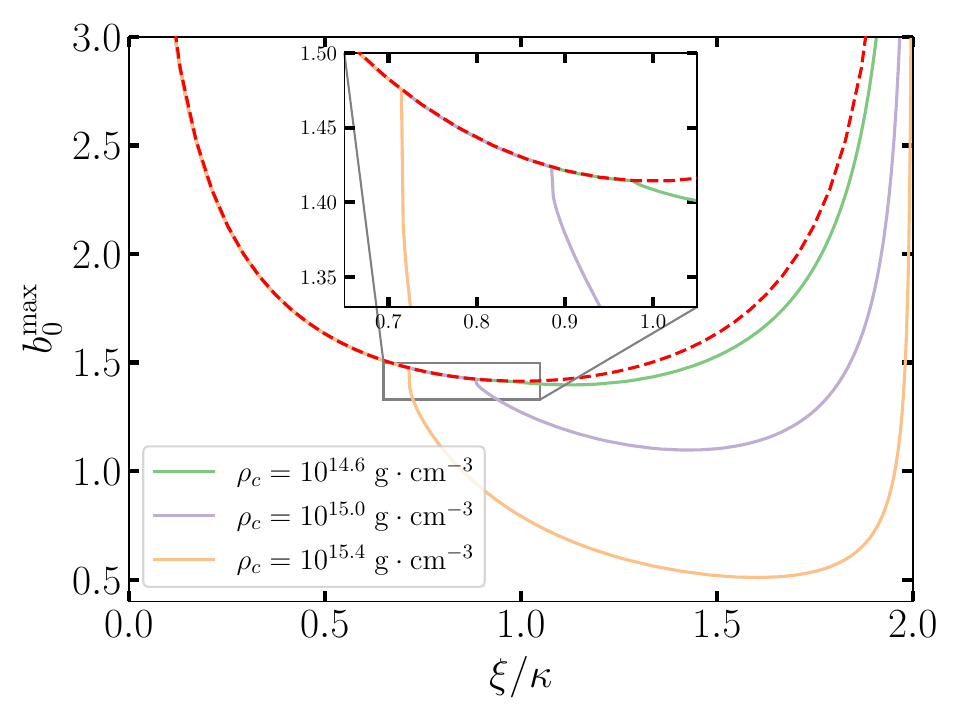}
\caption{\label{fig:b0max} The maximum value of $b_0$ at different central
densities when solving the modified TOV equations in the range $0<\xi<2\kappa$.
The red dashed line represents $p_2=0$ in Eq.~\eqref{eq:p2}. Here we set
$\nu_0=0$.}
\end{figure}

The parameter $b_0$ also has an upper bound within the range $0<\xi<2\kappa$, as
shown in Fig.~\ref{fig:b0max} for three different central densities. Above the
red dashed line, the pressure increases along the radial direction ($p_2>0$),
preventing being  a star.  However, the red line does not represent the
strictest bound as the coupling becomes stronger, as indicated by the curves
bifurcating from the red one.  The reason for this lies in the asymptotic
expansion equations of variables at infinity, discussed in
Appendix~\ref{appx:2}.  For the expansion to be valid, the coefficients must
remain finite, meaning that the denominator, $2\kappa+\xi(\xi-2\kappa)X^2$, must
be positive.  This requirement imposes a smaller maximum $b_0$ as $\xi$
increases.

\begin{figure}[t]
\includegraphics[width=\linewidth]{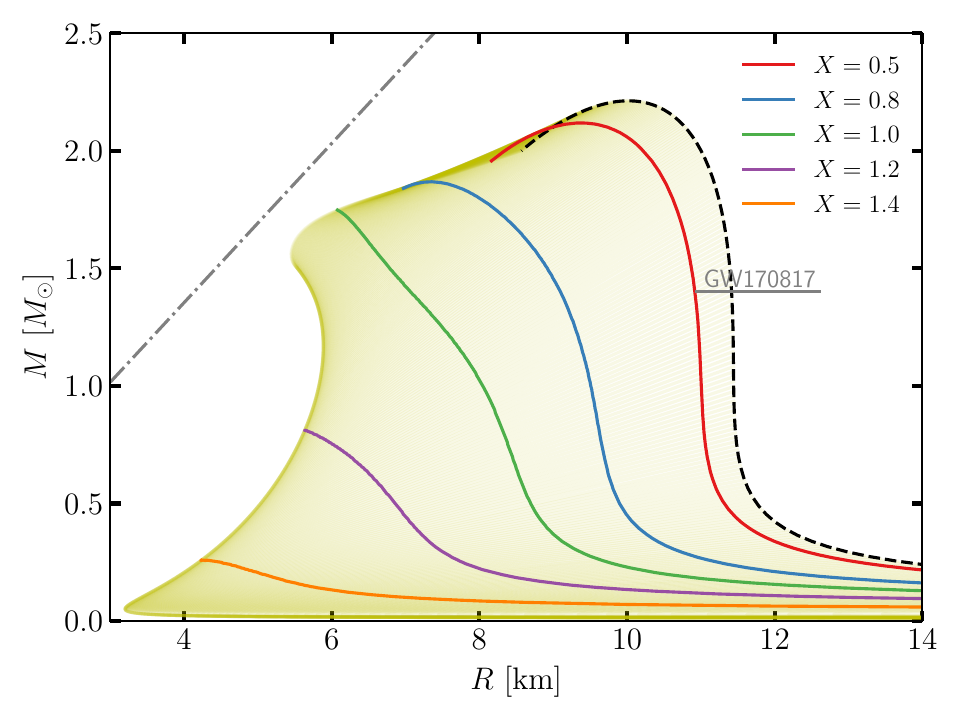}
\includegraphics[width=\linewidth]{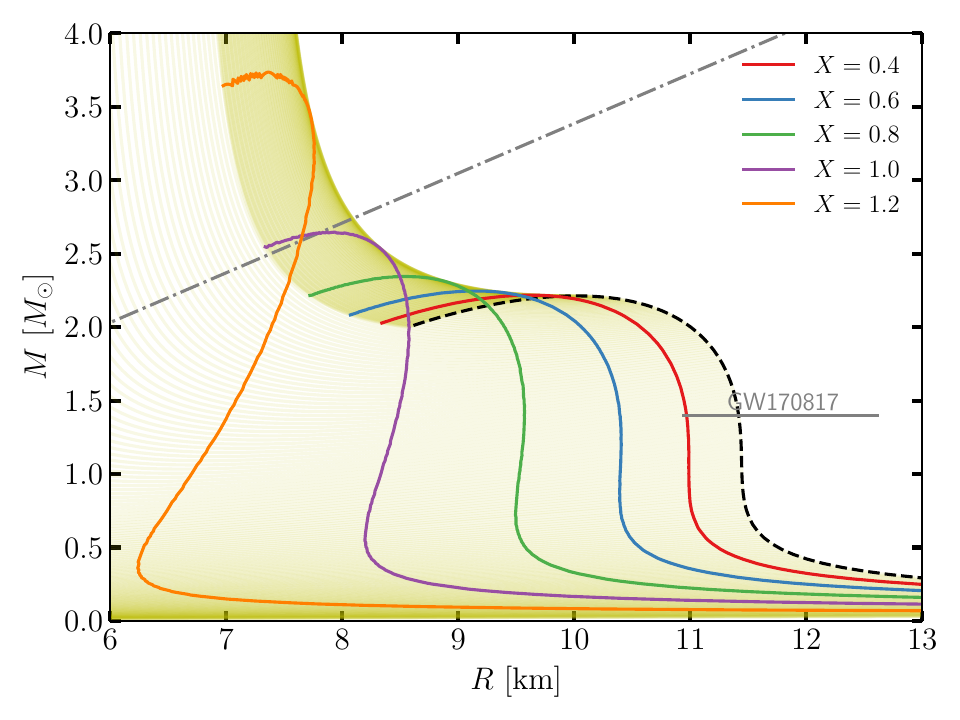}
\caption{\label{fig:MR} The mass-radius curves of NSs for $\xi=\kappa/2$ (top
panel) and $\xi=\kappa$ (bottom panel) in bumblebee theory with a same vector
field at infinity. The buff region, the black dashed line, the gray dot-dashed
line, and the short gray bar have the same meaning as in Fig.~\ref{fig:-1MR}.}
\end{figure}

To compare these two distinct cases, we plot the mass-radius relation for both
$\xi=\kappa/2$ and $\xi=\kappa$ in Fig.~\ref{fig:MR}.  For solutions where the
maximum value of $b_0$ is determined by $p_2$, the NS has a lower mass and a
smaller radius, as shown in the top panel.  However, when $b_0^\mathrm{max}$ is
constrained by the vector field at infinity, i.e. the right-hand side of the
bifurcation point, the NS can become a lot more massive than those in GR.  As
$b_0$ approaches its maximum value in this scenario, $X$ simultaneously
approaches the boundary where the denominator vanishes, leading to a dramatic
increase in the ADM mass and vector charge of the NS.  These extremely massive
NSs have smaller radii compared to others, resulting in smaller baryonic masses
and thus negative fractional binding energies.  Although such extremely massive
stars may not qualify as typical NSs due to their small radii, they remain
significant as massive objects since they do not give rise to any singularities.
They may constitute a new kind of exotic compact objects which are of central
focus in the GW astrophysics studies~\cite{Cardoso:2017cqb}.  Using the
aforementioned estimation of the NS radius, it is possible to constrain the
range of the vector field at infinity, leading to $X\lesssim0.40$ for
$\xi=\kappa/2$ and $X\lesssim0.42$ for $\xi=\kappa$.

\subsection{The $\xi>2\kappa$ case}

\begin{figure}[t]
\includegraphics[width=\linewidth]{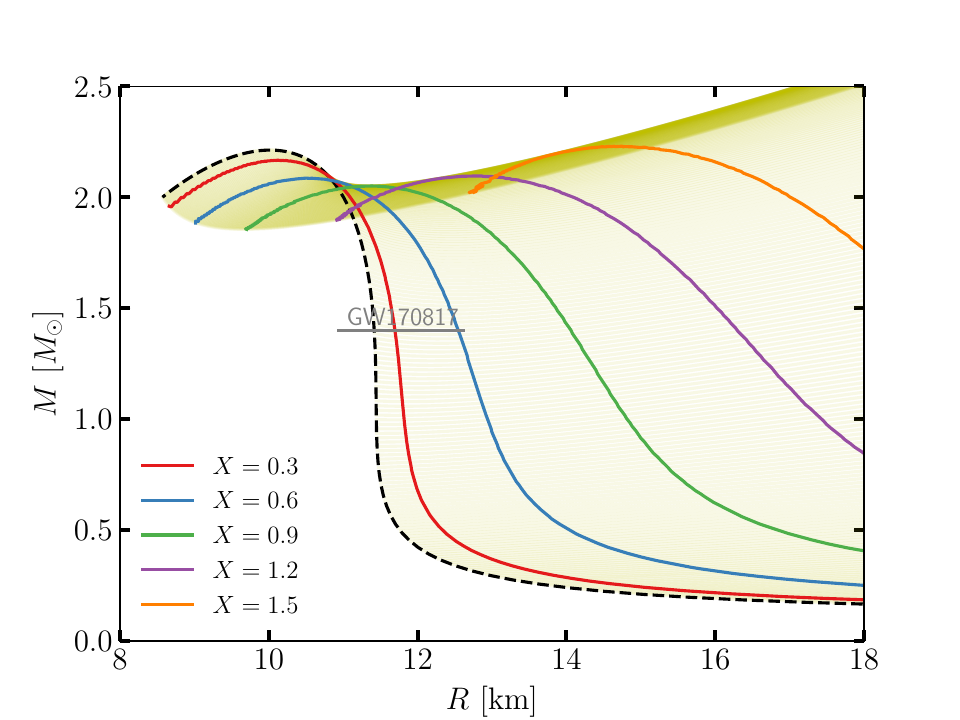}
\caption{\label{fig:3MR} The mass-radius curves of NSs for $\xi=3\kappa$ in
bumblebee theory with a same vector field at infinity. The buff region, the
black dashed line, and the short gray bar have the same meaning as in
Fig.~\ref{fig:-1MR}.}
\end{figure}

The mass-radius relation in the remaining part of the parameter space is
investigated for the representative case of $\xi=3\kappa$, with the result shown
in Fig.~\ref{fig:3MR}.  It can be seen that in this parameter space, the radii
of NSs in the bumblebee theory become larger.  If using the previous estimation
of radius from GW170817~\cite{Dietrich:2020efo} to impose constraints, we can
conclude that the time component of the asymptotic vector field cannot exceed
0.62.

Comparing Fig.~\ref{fig:-1MR} and Fig.~\ref{fig:3MR}, it is found that the buff
regions occupy similar positions in the mass-radius diagram, indicating some
commonalities between parameter spaces $\xi<0$ and $\xi>2\kappa$, as reflected
by the same constraints on parameters shown in Table~\ref{tab:paraspace}.
However, in the parameter space $\xi>2\kappa$, NSs can move far along the
positive direction of the radius and mass axes, and the corresponding solutions
are free of singularities.  This suggests that, in this case, extremely
high-mass and large-radius stars are allowed, with positive fractional binding
energy, unlike the massive stars in the parameter space $0<\xi<2\kappa$.

\section{Summary}\label{sec:sum}

The bumblebee theory of gravity is an important vector-tensor theory with local
Lorentz-symmetry violation.  For the first time, we have numerically calculated
the static spherical NS structure in this theory, exploring abundant
characteristics of NSs under various different couplings.

For the two specific coupling parameters $\xi=0$ and $\xi=2\kappa$, the
solutions obtained here are consistent with those in GR.  In the $\xi=0$ case,
the solution permits a trivial vector field, while in the $\xi=2\kappa$ case, a
non-trivial vector field is allowed while the exterior vacuum solution of the NS
corresponds to a portion of the Schwarzschild solution.  Using these two special
couplings as boundaries, we divided the range of the coupling parameter $\xi$
into three categories and discussed the characteristics of NS solutions for each
of them.

Vectorized NS solutions appear when $\xi<0$.  These stars carry a nonzero vector
charge and, as a result, can emit dipolar radiation when accelerated in
binaries.  The fractional banding energy of vectorized NSs is larger than the
corresponding value of NSs in GR, indicating that the vectorized stars are
energetically more favorable.  We also find that within this parameter range,
two different NSs can have exactly the same central density and central vector
field.  The only difference between them is the spacetime geometry, even though
both approach an asymptotically flat spacetime at infinity.  For certain values
of $\xi$ within the range $0<\xi<2\kappa$, there are NS solutions with a large
ADM mass but a small radius.  In contrast, for $\xi>2\kappa$, there are
solutions where both the ADM mass and the radius of NSs are large.  These
solutions are mathematically self-consistent and may potentially be used to
explain both observed and yet-to-be-detected special or exotic signals.  Except
for values of $\xi$ slightly greater than 0, all other couplings result in
scenarios where the stellar radius is smaller than the Schwarzschild radius
corresponding to its ADM mass, i.e. the compactness $C>0.5$.  In addition to
some theories of gravity that contain topological terms~\cite{Li:2023vbo}, NSs
within the bumblebee theory can also exhibit this special property with extreme
compactnesses.

Although the NS solutions in bumblebee theory exhibit a rich variety of
behaviors, not all of these can satisfy observational constraints.  We are the
first to use the estimation from GW170817 of stellar radius of a $M=1.4M_\odot$
NS to constrain parameters. Different maximum values of the vector field at
infinity were obtained for examples of different $\xi$ discussed in
Sec.~\ref{sec:num}.  We also used the observed maximum mass of
pulsars~\cite{Fonseca:2021wxt,Antoniadis:2013pzd} to constrain $X$ as well, but
this provides similar or less constraining limits, not to mention the large
uncertainties from NS EOSs.  Further constraints on the parameter space can be
obtained by perturbing the obtained NS solutions, including calculating the
tidal deformation and analyzing the stability of NSs, which could be the next
stage of investigation.  The effect of the self-interacting potential in
Eq.~\eqref{eq:potential} would emerge at the perturbation level, resulting in
deviations from the HN theory that might reveal new aspects of Lorentz-symmetry
violation.

In conclusion, as a generalization of the Einstein-Maxwell theory and an
important example of Lorentz-violating gravity theories, the bumblebee model
contains rich NS solutions.  We here have provided the first set of
comprehensive studies.  Testing these solutions plays a significant role in
fundamental physics.  It might help us explore whether gravity is nonminimally
coupled to a vector field similar to but beyond the electromagnetic field, and
it could also aid in investigating whether a special local reference frame
exists in the Universe.

\begin{acknowledgments}
We thank Zhanfeng Mai and Jinbo Yang for discussions.  This work was supported
by the National SKA Program of China (2020SKA0120300),  the National Natural
Science Foundation of China (11991053), the Beijing Natural Science Foundation
(1242018), the China Postdoctoral Science Foundation (2023M741999), the Max
Planck Partner Group Program funded by the Max Planck Society, and the 
High-performance Computing Platform of Peking University.
\end{acknowledgments}

\appendix

\begin{widetext}

\section{Field Equations Using the Static Spherical Ansatz}\label{appx:1}

Substituting ansatz \eqref{eq:ansatz} into the field equations, and assuming
that the bumblebee 1-form field has only time component $b_t$ and radial
component $b_r$.  The only two non-vanishing components of $\mathscr{E}^\mu$ are
\begin{eqnarray}
    &&\mathscr{E}^t=-\left\{b_t''-\left(\mu'+\nu'-\frac2r\right)b_t'+\left[\frac\xi\kappa\left(\mu'\nu'-\frac2r\nu'-\nu''-\nu'^2\right)-2V_\mathfrak{B}e^{2\mu}\right]b_t\right\}e^{-2(\mu+\nu)},\\
    &&\mathscr{E}^r=\left[\frac{\xi}{\kappa}\left(\mu'\nu'+\frac2r\mu'-\nu''-\nu'^2\right)-2V_\mathfrak{B}e^{2\mu}\right]b_re^{-4\mu}.
\end{eqnarray}
Here, $V_\mathfrak{B}=dV/d\mathfrak{B}={dV}/{d(B^\lambda B_\lambda)}$, which is
set to zero in the main text.  The nonzero components of $\mathscr{E}_{\mu\nu}$
are 
\begin{eqnarray}
    &&
    \begin{aligned}
        \mathscr{E}_{tt}
        &=\frac{e^{2\mu}+2r\mu'-1}{r^2}e^{2\nu-2\mu}-\kappa(V_\mathfrak{B}+\rho)e^{2\nu}-\left(\frac\kappa2-\xi\right)b_t'^2e^{-2\mu}\\
        &\quad+\xi\left\{b_t''-\left(\mu'+4\nu'-\frac2r\right)b_t'+\left[\left(2\mu'\nu'-\frac4r\nu'-2\nu''+\nu'^2\right)-\frac{2\kappa}\xi V_\mathfrak{B}e^{2\mu}\right]b_t\right\}b_te^{-2\mu}\\
        &\quad-\xi\left\{b_r'^2+\left[b_r''-\left(5\mu'-\frac4r\right)b_r'-\left(\mu''-3\mu'^2-\frac6r\mu'+\frac1{r^2}\right)b_r\right]b_r\right\}
    \end{aligned}\\
    &&
    \begin{aligned}
        \mathscr{E}_{rr}
        &=\frac{1+2r\nu'-e^{2\mu}}{r^2}+\kappa(V_\mathfrak{B}-p)e^{2\mu}+\left(\frac\kappa2b_t'^2-\xi b_tb_t'\nu'+\xi b_t^2\nu'^2\right)e^{-2\nu}\\
        &\quad+\xi\left[\left(\nu'+\frac2r\right)b_r'-\left(\nu''+\nu'^2-\frac2r\nu'-\frac1{r^2}\right)b_r\right]e^{-2\mu}-2\kappa V_\mathfrak{B}b_r^2,
    \end{aligned}\\
    &&
    \begin{aligned}
        \mathscr{E}_{\theta\theta}
        &=\left(\nu''+\nu'^2+\frac{\nu'-\mu'}r-\mu'\nu'\right)r^2e^{-2\mu}+\kappa r^2(V_\mathfrak{B}-p)-\left(\frac\kappa2b_t'^2-\xi b_tb_t'\nu'+\xi b_t^2\nu'^2\right)r^2e^{-2(\mu+\nu)}\\
        &\quad+\xi r^2\left\{b_r'^2+\left[b_r''-\left(5\mu'-2\nu'-\frac2r\right)b_r'+\left(\nu''-\mu''+\nu'^2-3\mu'\nu'+3\mu'^2+\frac{\nu'-3\mu'}{r}\right)b_r\right]b_r\right\}e^{-4\mu},
    \end{aligned}\\
    &&\mathscr{E}_{\phi\phi}=\sin^2\theta\mathscr{E}_{\theta\theta},\\
    &&\mathscr{E}_{tr}=\mathscr{E}_{rt}=\kappa e^{2\mu}\mathscr{E}^rb_t=\kappa\mathscr{E}_rb_t.
\end{eqnarray}

\section{Asymptotic Expansions of Variables at Infinity}\label{appx:2}

In fact, the equations of motion restrict $g_{rr}=1$ at infinity but leave the
asymptotic value of $g_{tt}$ free, which is denoted by $\nu_\infty$.  Comparing
equations of motion order by order gives the recurrence relations of
coefficients in Eq.~\eqref{eq:infexpand} as follows
\begin{eqnarray}
    \nu_{-1}&=&-GM,\\
    m_{-1}&=&\frac{16\pi^2\kappa(\kappa-2\xi)\alpha^2+6\sqrt{\pi\kappa}\kappa\xi\alpha X+3\kappa G\xi X^2+\xi(\xi-2\kappa)[8\pi^2\kappa\alpha^2+\frac{\xi}{2}(GX+\sqrt{\pi\kappa}\alpha)X]X^2e^{-2\nu_\infty}}{2e^{2\nu_\infty}\kappa+\xi(\xi-2\kappa)X^2}M^2,\quad \quad \quad\\
    \nu_{-2}&=&-\frac{2\pi\kappa^2(\xi-\kappa)\alpha^2+2\sqrt{\pi\kappa}\kappa G\xi\alpha X+G^2(2e^{2\nu_\infty}\kappa+\xi(\xi-\kappa)X^2)}{2e^{2\nu_\infty}\kappa+\xi(\xi-2\kappa)X^2}M^2,\\
    b_{-2}&=&-\frac{2\pi\kappa(\xi-\kappa)\alpha^2+2\sqrt{\pi\kappa}G\xi\alpha X+G^2\xi X^2}{2e^{2\nu_\infty}\kappa+\xi(\xi-2\kappa)X^2}\xi M^2X, \\
    & \cdots &
\end{eqnarray}
Here, $\alpha$ is the dimensionless charge-to-mass ratio defined as
$\alpha\equiv {Q}/{\sqrt{8\pi G}M}$.  The assumption of asymptotic flatness
selects $\nu_\infty=0$.

There are three free variables---$M$, $X$, and $Q$---that determine the vacuum
solution.  However, only two of these three variables are independent, both for
BH solutions and NS solutions.  For a spherically symmetric static BH in the
bumblebee theory, $g_{tt}$ must be zero at the horizon where $g^{rr}$ vanishes,
leading to a constriant $f^\mathrm{BH}(\xi;M,Q,X)=0$~\cite{Xu:2022frb}.  In the
case of a spherically symmetric, static NS, all variables are determined by the
central density $\rho_c$ and the central vector field $b_c$ for a given EOS.
Therefore, a similar constraint $f^\mathrm{NS}(\mathrm{EOS},\xi;M,Q,X)=0$ must
exist, indicating that the relation between the ADM mass, vector charge, and the
background field depends on the coupling $\xi$ as well as the specific EOS.  If
the expressions of $f^\mathrm{BH}$ and $f^\mathrm{NS}$, as functions of $M$, $Q$
and $X$, are identical for any EOS of NSs and the same $\xi$, then we can
conclude that an extended version of Birkhoff theorem holds in the bumblebee
theory with this specific coupling.

Our numerical calculations reveal that, for any EOS and any boundary conditions
at the center, the spacetime outside the NS is the Schwarzschild spacetime when
$\xi=2\kappa$.  Previous numerical studies on BHs in this theory also found that
all spherically symmetric, static BH solutions are stealth Schwarzschild,
meaning that the spacetime is Schwarzschild with a nonzero $b_t(r)$ when
$\xi=2\kappa$.  This result supports the implication that an extended form of
the Birkhoff theorem holds in this theory, at least for $\xi=2\kappa$.

\section{Another Form of Modified TOV Equations}\label{appx:3}

It is easy to notice that $\mu$ and $\mu'$ can be eliminated from the other
variables by using the field equations. The expressions of $\mu$ and $\mu'$ are
\begin{eqnarray}
    \mu&=&\frac{1}{2}\ln\left[\frac{\mathscr Fr^2+2e^{2\nu}(1+2r\nu')}{2(1+\mathfrak p)}\right]-\nu,\\
    \mu'&=&\frac{2\nu'+2[(1+\mathfrak p)(\nu''+\nu'^2)r-\mathfrak p(1+r\nu')r^{-1}]-e^{-2\nu}(1+2\mathfrak p)\mathscr Fr}{2(1+\mathfrak p)(1+r\nu')},
\end{eqnarray}
where $\mathfrak p=pr^2$, $\varrho=\rho r^2$ and
\begin{equation}
	\mathscr{F}= b_t'^2-2\xi  b_t b_t'\nu'+2\xi b_t^2\nu'^2.
\end{equation}

\noindent Then the equations of $\nu$ and $b_t$ become
\begin{eqnarray}\label{eq:b4}
    \nu''=\frac{\mathscr V}{\mathscr{N}},\quad b_t''=\frac{\mathscr X}{\mathscr N},
\end{eqnarray}
where
\begin{eqnarray}
    \mathscr V&=&e^{-2\nu}(2-\xi)\xi b_t^2\nu'\mathscr Fr^2(1+2\mathfrak p)+2e^{2\nu} \Big\{(1+r\nu')\big[\varrho(1+2r\nu')r^{-1}-4\nu'\big]+\big[3+r\nu'(3-2r\nu')\big]\mathfrak pr^{-1} \Big\}\nonumber\\
	&&+\;r b_t^2\Big\{2+(\varrho-2\xi)(1+r\nu')+\big[5+r\nu'-2\xi(1+r\nu')\big]\mathfrak p \Big\}+2\xi b_t b_t'\nu' \big[2(1+2r\nu')-\varrho(1+r\nu')-(1-3r\nu')\mathfrak p \big]r\nonumber\\
	&&+\;2\xi b_t^2\nu' \Big\{(2+3\mathfrak p)(2-\xi)+r\nu' \big[2(1-\xi-2r\nu')+\varrho(1+r\nu')+(9-4\xi-3r\nu')\mathfrak p \big] \Big\},\\
    \mathscr X&=&e^{-2\nu}(2-\xi)\xi b_t^2 b_t'\mathscr Fr^2(1+2\mathfrak p)+2e^{2\nu} \Big\{ \big[\varrho b_t'+(\varrho+3\mathfrak p)\xi b_tr^{-1}\big](1+2r\nu')- \big[4+(3-2r\nu')\mathfrak p\big] b_t' \Big\}\nonumber\\
	&&+\; b_t'^3r^2 \big[\varrho+\mathfrak p-2\xi(1+\mathfrak p) \big]+2\xi^2 b_t^3r\nu'^2(\varrho+\mathfrak p-2)+\xi b_t b_t'^2r \big[2+\varrho+5\mathfrak p-2\xi(1+\mathfrak p)+2(4-\varrho+3\mathfrak p)r\nu' \big]\nonumber\\
	&&+\;2\xi b_t^2 b_t' \Big\{(2+3\mathfrak p)(2-\xi)+ \big[\xi(2-\varrho)+(\varrho-4)r\nu'+(4-3\xi-3r\nu')\mathfrak p \big]r\nu' \Big\},\\
    \mathscr N&=&2(1+\mathfrak p)r\mathscr C=2(1+\mathfrak p) \big[2e^{2\nu}+\xi(\xi-2) b_t^2 \big]r.
\end{eqnarray}
The denominator $\mathscr{N}$ linearly approaches zero when approaching the
stellar center.  To ensure that the equation is not singular at $r=0$, the
numerator must approach zero linearly or faster.  A term-by-term analysis of
$\mathscr{V}$ and $\mathscr{X}$ ultimately leads to the requirement that
$\nu'=b_t'=0$ at $r=0$.

\section{Differential Equations Outside a NS}\label{appx:4}

Since $\rho=p=0$ outside the star, we only need to solve metric functions $\nu$,
$\mu$ and the vector field $b_t$.  The differential equations of $\nu(x)$ and
$b_t(x)$ are
\begin{eqnarray}
    \ddot\nu=\frac{\mathcal V}{\mathcal N},\quad \ddot{ b}_t=\frac{\mathcal X}{\mathcal N},
\end{eqnarray}
where the overdot represents the derivative with respect to $x$.  With the
following definitions,
\begin{eqnarray}
    \mathcal F(x)&=&\dot{ b}_t^2-2\xi b_t\dot{ b}_t\dot\nu+2\xi b_t^2\dot\nu^2,\\
	\mathcal C(x)&=&2e^{2\nu}+\xi(\xi-2) b_t^2,\\
	\mathcal N(x)&=&2(1+\beta)(1+\beta x)\mathcal C(x),\\
	\mathcal H(x)&=&(1-x)(1+\beta x)\dot\nu,
\end{eqnarray}
$\mathcal{V}$ and $\mathcal{X}$ are given by
\begin{eqnarray}
    \mathcal V&=&(1+\beta x)\dot{ b}_t^2 \Big\{2(1+\beta)(1-\xi)-\xi \big[2e^{2\nu}+(\xi-2) b_t^2 \big]\mathcal He^{-2\nu} \Big\}-8e^{2\nu}(1+\beta) \big[\beta+(1+\beta x)\dot\nu \big]\dot\nu\nonumber\\
	&&+\;2(1+\beta x)\xi^2(2-\xi) b_t^4\dot\nu^2\mathcal He^{-2\nu}+2(1+\beta x)\xi b_t\dot{ b}_t\dot\nu \big[2(1+\beta)+(2+\mathcal Ce^{-2\nu})\mathcal H \big]\nonumber\\
    &&-\;4\xi b_t^2 \Big\{\beta(1+\beta)(\xi-2)+(1+\beta x) \big[(1+\beta)(\xi-1)+2\mathcal H \big]\dot\nu \Big\}\dot\nu,\\
    \mathcal X&=&-(1+\beta x)\xi \big[2e^{2\nu}+(\xi-2) b_t^2 \big]\dot\nu^2\mathcal He^{-2\nu}-4(1+\beta)(1+\beta x)^2\xi^2 b_t^3\dot\nu^2\nonumber\\
	&&+\;2(1+\beta x)\xi b_t\dot{ b}_t^2 \big[(1+\beta)(1-\xi)+(2+\mathcal Ce^{-2\nu})\mathcal H \big]-8e^{2\nu}\beta(1+\beta)\dot{ b}_t\nonumber\\
	&&-\;2\xi b_t^2 \Big\{2\beta(1+\beta)(\xi-2)+(1+\beta x) \big[(2+\mathcal Ce^{-2\nu})\mathcal H-2(1+\beta)\xi\big]\dot\nu \Big\}\dot{ b}_t.
\end{eqnarray}
The expression of $\mu$ and $\dot\mu$ are then derived as
\begin{eqnarray}
    \mu&=&-\nu+\frac12\ln\left[\frac{1+\beta+2(1-x)(1+\beta x)\dot\nu}{1+\beta}e^{2\nu}+\frac{(1-x)^2(1+\beta x)^2}{2(1+\beta)^2}\mathcal F\right],\\
    \dot\mu&=&-\dot\nu+\frac{(1-x)(1+\beta x)\xi}{2(1+\beta)\mathcal C} \big[(\xi-2) b_t^2\mathcal Fe^{-2\nu}-2(\dot{ b}_t-2 b_t\dot\nu)^2 \big].
\end{eqnarray}

\begin{figure*}[t]
\includegraphics[width=\linewidth]{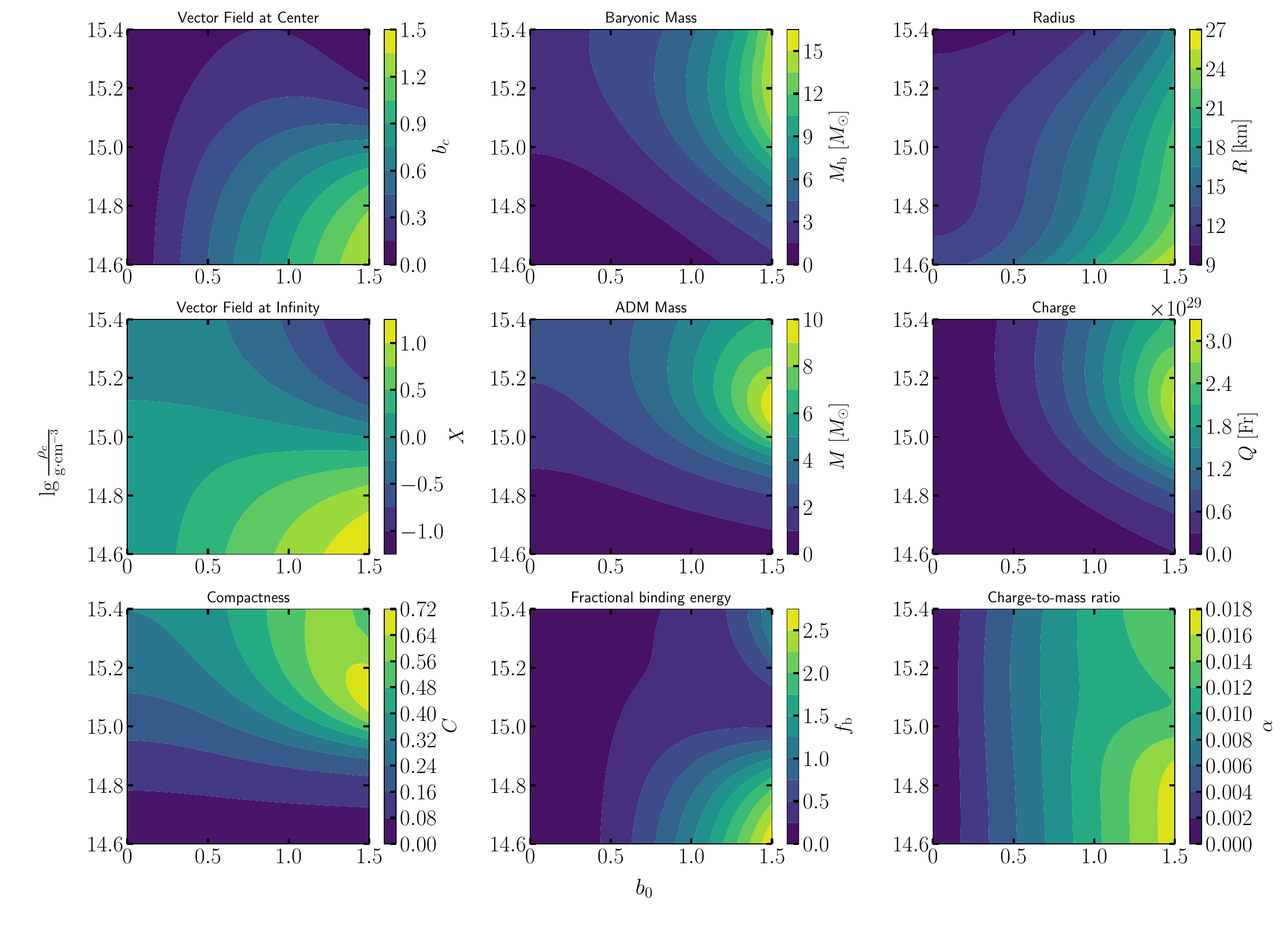}
\caption{\label{fig:appx:1} Contour plots for nine quantities  on the
$b_0$-$\rho_c$ plane of NS solutions for a nonminimal coupling $\xi=-\kappa$:
the vector field at center and infinity, baryonic and ADM masses, radius and
charge of NSs, as well as the compactness, fractional binding energy, and
charge-to-mass ration.  The unit of the vector charge is Franklin
($1\;\mathrm{Fr}=1\;\mathrm{cm^{3/2}\cdot g^{1/2}\cdot s^{-1}}$) in Gaussian
units.  The largest $b_0$ in horizontal axis does not reach its theoretical
maximum $b_0^\mathrm{max}$, and it is truncated at $b_0=1.5$ to avoid stellar
radii that are too large in the plot.}
\end{figure*}
\begin{figure*}[t]
\includegraphics[width=\linewidth]{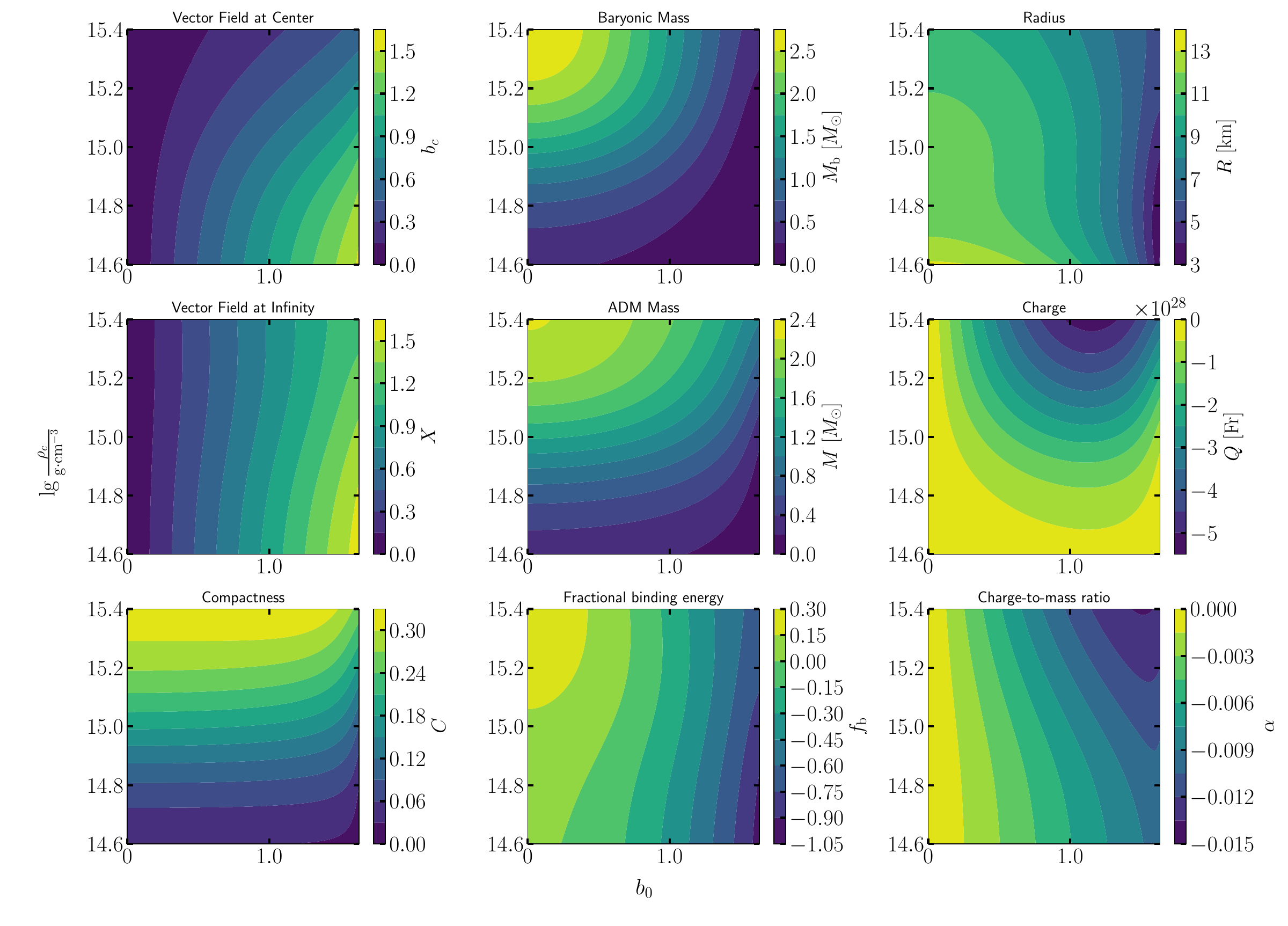}
\caption{\label{fig:appx:2} Same as Fig.~\ref{fig:appx:1}, but for
$\xi=\kappa/2$.  The largest value of the horizontal axis is $\sqrt{2}$, which
is the maximum of $b_0$ in this case.}
\end{figure*}
\begin{figure*}[t]
\includegraphics[width=\linewidth]{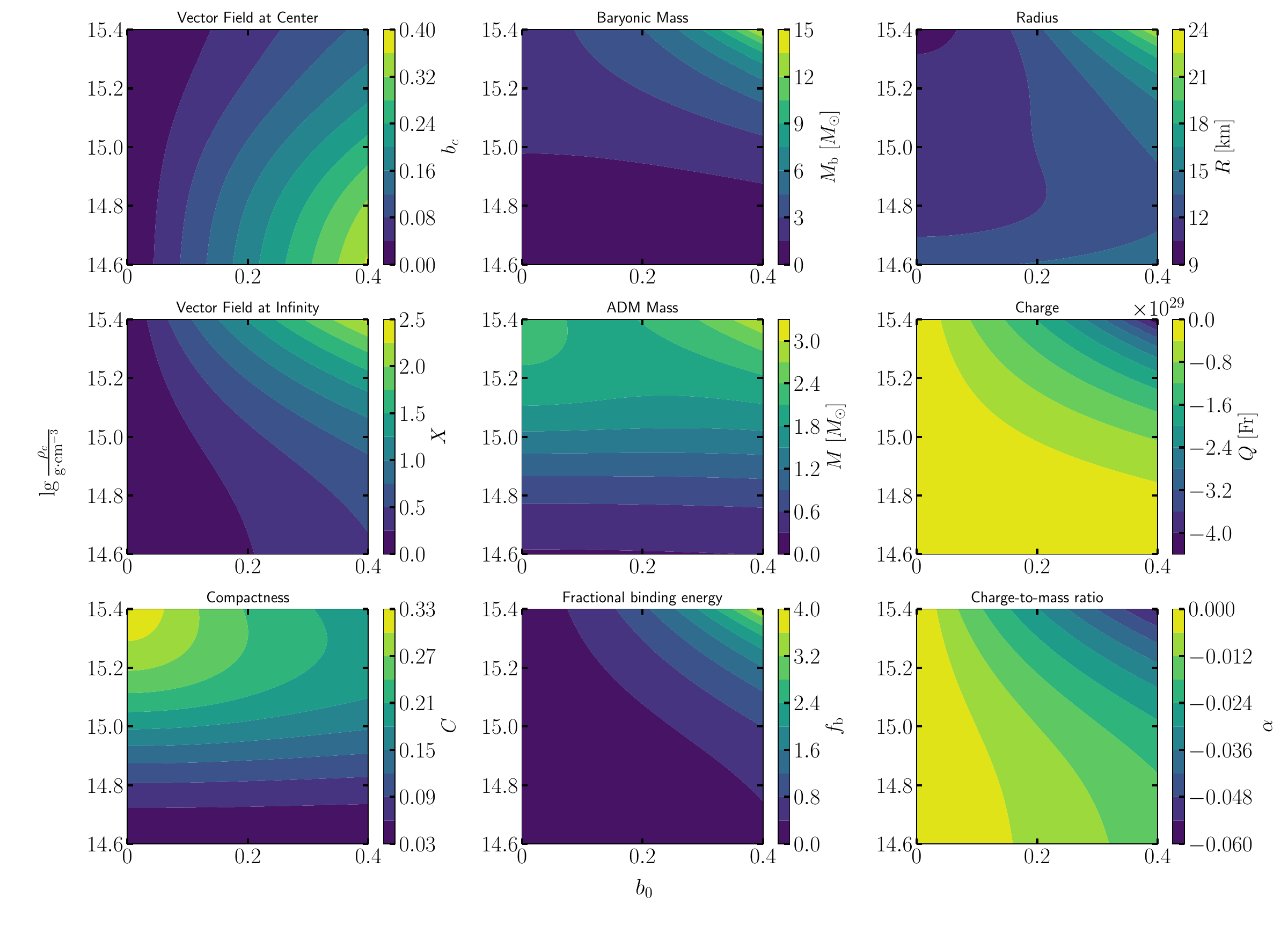}
\caption{\label{fig:appx:3} Same as Fig.~\ref{fig:appx:1}, but for
$\xi=3\kappa$. The horizontal axis is truncated at $b_0=0.4$ to avoid stellar
radii that are too large in the plot.}
\end{figure*}

\section{Supplementary Results}\label{appx:5}

Here we present three contour plots in Figs.~\ref{fig:appx:1}--\ref{fig:appx:3}
for the time component of the vector field at the center $b_c$, baryonic mass
$M_\mathrm{b}$, stellar radius $R$, time component of the vector field at
infinfty $X$, ADM mass $M$, vector charge $Q$, compactness $C$, fractional
binding energy $f_\mathrm{b}$, and charge-to-mass ratio $\alpha$ on the
$b_0$-$\rho_c$ plane.  The three figures respectively depict the cases for
$\xi=-\kappa$, $\xi=\kappa/2$, and $\xi=3\kappa$, with $\nu_0=0$ taken for the
solutions.  The reason the case $\xi=\kappa$ is not plotted is that, in this
scenario, the mass and charge diverge in certain regions of the parameter space.
We include these figures in this appendix as supplementary material to help
readers look deeper into the NS solutions.  Additionally, if future observations
can place tight constraints on certain parameters, like the compactness or the
charge-to-mass ratio, these figures could provide a helpful guidance.

\end{widetext}


\bibliography{bumNS}

\end{document}